\documentclass[twocolumn,aps,prx,showpacs,superscriptaddress,groupedaddress]{revtex4}
\usepackage{textcomp}
\usepackage{amsmath}
\usepackage{amssymb}
\usepackage{graphicx}
\usepackage{color}

\begin{document}

\title{Free space quantum communication with a portable quantum memory}
\author{Mehdi Namazi}
\affiliation{Department of Physics and Astronomy, Stony Brook University, New York 11794-3800, USA}
\author{Giuseppe Vallone}
\affiliation{Department of Information Engineering, University of Padova, Via Gradenigo 6b, 35131 Padova, Italy}
\author{Bertus Jordaan}
\author{Connor Goham}
\author{Reihaneh Shahrokhshahi}
\affiliation{Department of Physics and Astronomy, Stony Brook University, New York 11794-3800, USA}
\author{Paolo Villoresi}
\affiliation{Department of Information Engineering, University of Padova, Via Gradenigo 6b, 35131 Padova, Italy}
\author{Eden Figueroa}
\affiliation{Department of Physics and Astronomy, Stony Brook University, New York 11794-3800, USA}

\begin{abstract}
The realization of an elementary quantum network that is intrinsically secure and operates over long distances requires the interconnection of several quantum modules performing different tasks. In this work we report the interconnection of four different quantum modules: (i) a random polarization qubit generator, (ii) a free-space quantum communication channel, (iii) an ultra-low noise portable quantum memory and (iv) a qubit decoder, in a functional elementary quantum network possessing all capabilities needed for quantum information distribution protocols. We create weak coherent pulses at the single photon level encoding polarization states $|H\rangle, |V\rangle, |D\rangle, |A\rangle$ in a randomized sequence. The random qubits are sent over a free-space link and coupled into a dual rail room temperature quantum memory and after storage and retrieval are analyzed in a four detector polarization analysis akin to the requirements of the BB84 protocol. We also show ultra-low noise and fully-portable operation, paving the way towards memory assisted all-environment free space quantum cryptographic networks.
\end{abstract}

\pacs{03.67.Hk, 42.50.Ct, 42.50.Gy}

\maketitle

\section{Introduction}
The field of quantum information has recently seen remarkable progress regarding the implementation of elementary quantum devices and quantum communication protocols. On one hand, the advent of photonic quantum communication using long distance free space links \cite{capraro_impact_2012,vallone_free-space_2014,vallone_experimental_2015,schm07prl,naue13npho} has opened the possibilities to securely exchange quantum states and entanglement \cite{ritter_elementary_2012,choi_entanglement_2010,nolleke_efficient_2013}. These developments together with quantum key distribution protocols have enormous potential for the creation of a global, secure quantum information exchange network \cite{scarani_security_2009,MDI_QKD_2012,bacco_experimental_2013,liu_experimental_2013,tang_experimental_2014,abruzzo_measurement-device-independent_2014,panayi_memory-assisted_2014}.
\begin{figure*}
\centerline{\includegraphics[width=1.8\columnwidth]{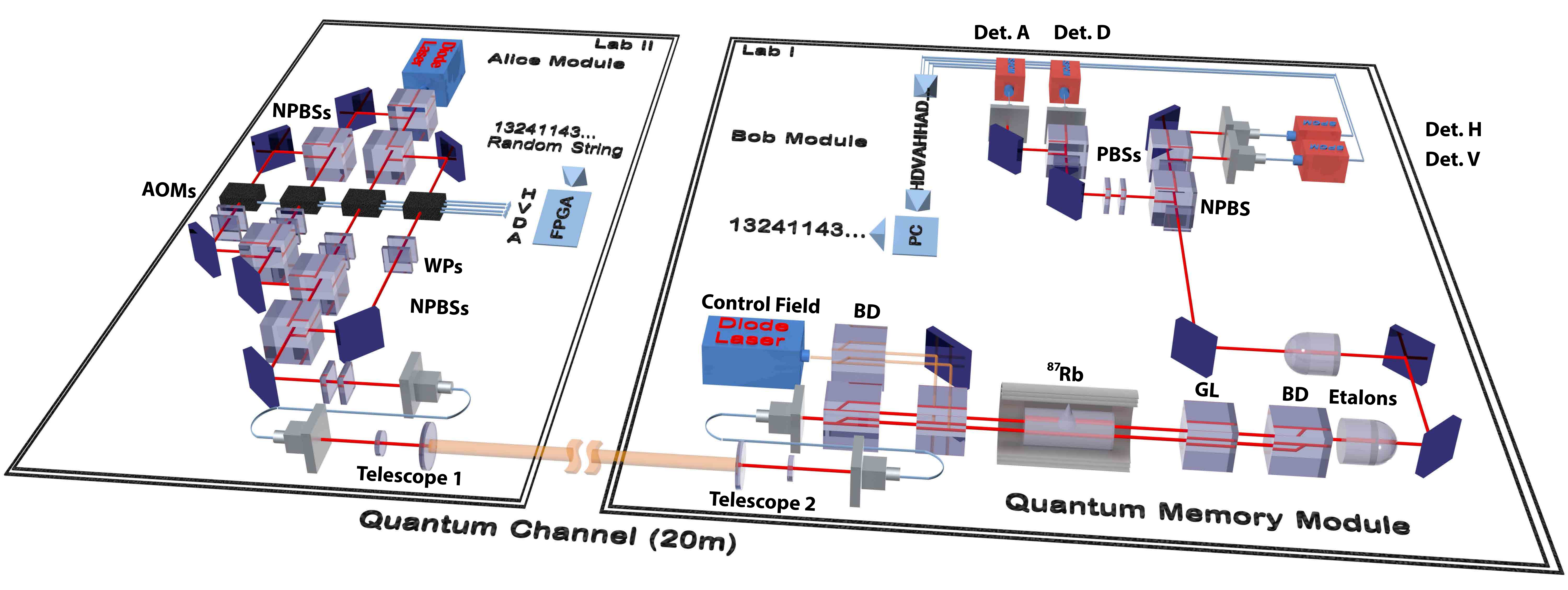}}
\caption{\textbf{Experimental setup for free-space quantum communication.} In Laboratory II Alice creates a random sequence of four orthogonal qubits ($|H\rangle, |V\rangle, |D\rangle, |A\rangle$). The 400ns-long qubits are produced every 40 $\mu$s. The qubits propagate in a free-space quantum communication channel over a distance of $\sim$ 20m and are then directed into a dual rail room temperature rubidium vapor quantum memory in Laboratory I. The control storage pulses are time-optimized to the arrival of the qubits in front of the memory. In Bob's site a four detector setup measures all possible basis at the exit of the memory to determine the quantum bit error rate (QBER). PBS: polarizing beam splitter, WP: wave plates, AOM: acousto-optical modulator, BD:beam displacer, GL: Glan-laser polarizer.}
\end{figure*}
On the other hand, an entirely different community of quantum scientists has developed sophisticated quantum light matter interfaces capable of receiving, storing and retrieving photonic qubits \cite{reiserer_cavity-based_2015,northup_quantum_2014,bussieres_prospective_2013,heshami_quantum_2015}. Such devices, collectively known as quantum memories already operate with high fidelities\cite{riedl_bose-einstein_2012,gundogan_quantum_2012,saglamyurek_quantum_2015}, long storage times \cite{novikova_electromagnetically_2012,hsiao_eit-based_2016} and high storage efficiencies \cite{hosseini_high_2011,yang_efficient_2016}. Furthermore, quantum memories already operate at room temperature \cite{namazi_unconditional_2015,england_storage_2015,saunders_cavity-enhanced_2016}, thus facilitating their interconnection with other quantum devices.

The construction of an interconnected set of many quantum devices that performs secure communication protocols in outside settings and with moving targets it is now within experimental reach \cite{bouwmeester_physics_2000,cirac_quantum_1997,kimble_quantum_2008}. Therefore it is of utmost relevance to engineer elementary networks of a few quantum nodes and quantum channels in order to understand the potential of these novel architectures \cite{datta_compact_2012,komar_quantum_2014,Reiserer_PRX_2016,Felix_2014}. The emergent behaviour of such small quantum networks should allow us to realize more sophisticated quantum procedures \cite{duan_long-distance_2001}. An important example of such an elementary network will be the modular connection of quantum cryptography systems operating over free-space quantum channels \cite{vallone_interference_2016}, assisted by room temperature quantum memories increasing the distance, security and connectivity of quantum key distribution protocols \cite{panayi_memory-assisted_2014,abruzzo_measurement-device-independent_2014}.\\
Paramount to the creation of such a free-space memory assisted quantum communication network is the use of shot-by-shot unconditional quantum memories capable of supporting the specific technical demands of outside-of-the-laboratory quantum communication channels. Among them, accepting random qubit states necessary to perform quantum key distribution protocols, having a minimized quantum bit error rate (QBER) and receiving spatially multi-mode signals, while simultaneously being cost effective and fully-portable. These capabilities will allow the construction of elementary quantum networks without the need for frequency conversion among their components, that are intrinsically secure, quantum coherent and compatible with long distance operation.\\
Here we report the creation of such an elementary quantum network in which we mimic these desired properties into a scaled down setup. We present individual experiments addressing the various challenges in order to create the quantum connectivity needed to perform memory-assisted long distance communication of random polarization qubits. To our knowledge, our results represent the first time that the ideas of quantum communication, as used in the well known BB84 protocol, are combined with low-noise room-temperature quantum storage. Our results are obtained by cascading four different quantum modules: a random polarization qubit generator, a free space quantum communication channel,warm vapor quantum memory and a qubit decoder.

\section{Experimental procedure.}
\subsection{Preparation of a random stream of qubits: Alice Module.}
Our elementary quantum network starts with the creation of a sequence of four polarization states ($|H\rangle, |V\rangle, |D\rangle=1/\sqrt 2(|H\rangle+|V\rangle), |A\rangle=1/\sqrt 2(|H\rangle-|V\rangle)$) in a distant laboratory (Alice's station, Laboratory II in Fig. 1). We create the qubits using 400ns-long pulses produced every 40 $\mu s$  by 4 individual acousto-optical modulators (AOMs). In order to compensate for small deviations in the length of each AOM track, the AOMs are each driven by independent sources regarding their amplitude and frequency modulation. The setup is designed to generate either an ordered sequence of four qubits in cycles of 160 $\mu s$ (see Fig. 2) or a train of qubit pulses where the modulation sources are controlled by a FPGA chip programmed to randomly trigger one of the four AOM's. The resulting random sequence of pulses is attenuated to the single-photon-level and then sent into free space quantum channel module.
\subsection{Propagation of qubit streams: Free space quantum channel module.}
The qubits created in the Alice station propagate in a free-space quantum communication channel over a distance of $\sim$ 20m without shielding or vacuum propagation and are then directed to a quantum memory setup in a different laboratory. We have chosen the characteristics of this setup as a test bed of the interconnectivity of this station and the quantum memory setup under more challenging out-of-the-laboratory operation. Of particular interest are the shot-by-shot changes in the mean input photon number due to the air turbulence between the laboratories and the capability of the memory to receive random polarization inputs, pulse-by-pulse. By careful alignment the loss in the free space propagation is set to be less than 4\%. Together with 63\% fiber coupling efficiency at the receiving end of the quantum memory setup this yields a total transmission of 59\% for the quantum communication channel. The shot-by-shot fluctuations in the mean photon number were measured to be $\sim$ 5\%.
\begin{figure*}
\includegraphics[width=2.0\columnwidth]{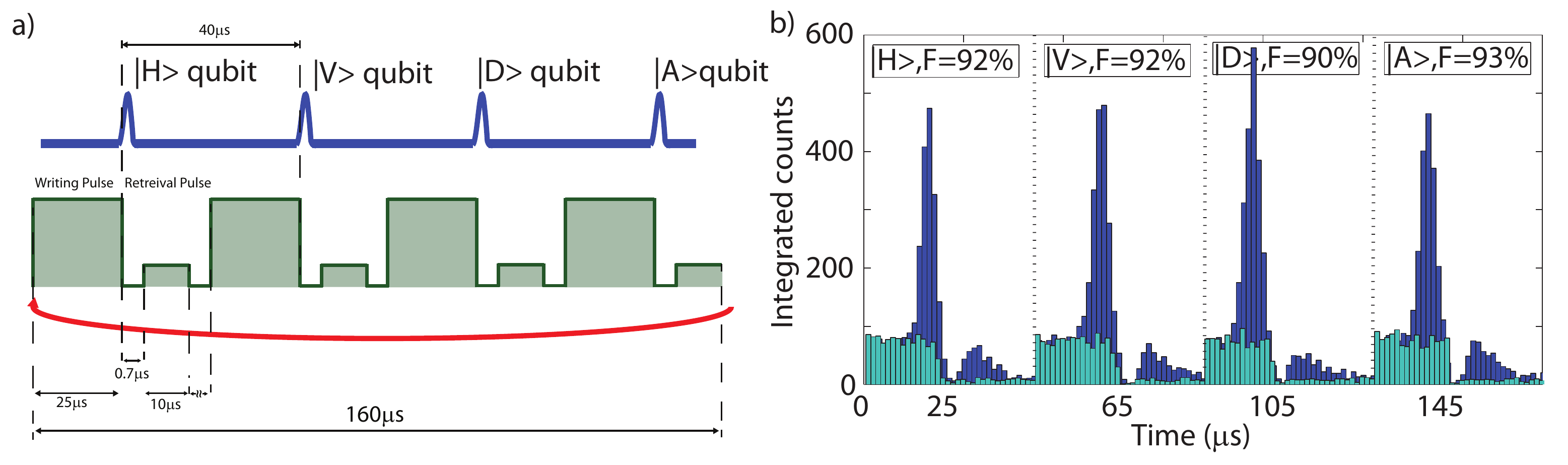}
\caption{\textbf{Storage of a sequence of qubits.} (a) A stream of polarization qubits with on average 3.5 photons propagates through a free space quantum communication channel of 20m. In the quantum memory site, the single-photon level qubits are received and stored sequentially using timed control field pulses. (b) Histograms for each of the polarization inputs after storage (dark blue) and background floor (light blue). Each histogram is presented in a 2$\mu$s time interval (see dashed black divisions). The fidelities are estimated from the signal-to-background ratio}.
\end{figure*}
\subsection{Storage of incoming pulses: Quantum Memory Module}
Located in Laboratory I is the room temperature quantum memory in which we store the incoming qubits. The quantum memory is based upon a warm $^{87}$Rb vapor and controlled using electromagnetically induced transparency (EIT). Two independent control beams coherently prepare two volumes within a single $^{87}$Rb vapor cell at $60^{\circ}\,$C, containing Kr buffer gas to serve as the storage medium for each mode of the polarization qubit. We employed two external-cavity diode lasers phase-locked at 6.835 GHz. The probe field frequency is stabilized to the $5S_{1/2} F = 1$ $\rightarrow$ $5P_{1/2} F' = 1$ transition at a wavelength of 795 nm while the control field interacts with the $5S_{1/2} F = 2$ $\rightarrow$ $5P_{1/2} F' = 1$ transition. Polarization elements supply 42 dB of control field attenuation (80\% probe transmission) while two temperature-controlled etalon resonators (linewidths of 40 and 24 MHz) provide additional 102 dB of control field extinction. The total probe field transmission is 4.5\% for all polarization inputs, exhibiting an effective, control/probe suppression ratio of 130 dB \cite{namazi_unconditional_2015}. The control field pulses are time-optimized to the arrival of the qubits in front of the memory (see Fig. 2a).

\subsection{Measuring the random stream of qubits: Bob module.}
After passing through the polarization independent frequency filtering system, the stored pulses enter the Bob module, which is equipped with a non-polarizing beam splitter (separating the $Z=\{|H\rangle, |V\rangle\}$ and $X=\{|D\rangle, |A\rangle\}$ bases) and two polarizing beam splitters whose outputs are detected by four single-photon counting modules (SPCM). Each SPCM corresponds to a different polarization state. This allow us to compare the detected sequence with the originally sent qubits and estimate the influence of the photonic background of the memory in the evaluation of the QBERs.

\section{Experiment 1: Storage of a sequence of four polarization qubits after free space propagation.}
In our first experiment, a string of four ordered  polarization qubits ($|H\rangle$, $|V\rangle$, $|D\rangle$ and $|A\rangle$) is sent from Alice module to the memory and Bob terminal through the free space channel in order to test the compatibility of all the modules and the performance of the quantum memory at the single photon level (see Fig. 2). The characterizations of the qubits after storage is done with a single detector placed after the memory bypassing the polarization analysis setup. We create histograms using the time of arrival and estimate a best-case-scenario fidelity of the stored polarization qubits containing on average 1.6 photons per pulse right before the memory. \\
We evaluate the signal to background ratio (SBR) in the measurements, defined as $\eta/q$, where $\eta$ is the retrieved fraction of a single excitation stored in a quantum memory and $q$ the average number of concurrently emitted photons due to background processes. Both are calculated by integrating the retrieved and background signals over 100 ns intervals. The Fidelities are then estimated as $F=1-\frac{1}{2}\frac{q}{\eta}$. Our analysis shows that even with the additional constraint of shot-by-shot fluctuations in intensity due to free space propagation and the addition of randomly polarized background photons in the memory, maximum fidelities of 92\% for $|H\rangle$, 92\% for $|V\rangle$, 90\% for $|D\rangle$ and 93\% for $|A\rangle$ can still be achieved. \\
These results are clearly above the classical threshold limit of 85\% for the corresponding efficiencies thus providing the necessary condition of unconditional quantum memory operation \cite{namazi_unconditional_2015}. They also show that our room temperature quantum memory implementation operates with the same parameters regardless of the polarization input, a fundamental attribute if the memory were to work as either a synchronization device for quantum cryptography protocols in which a stream of random qubits is used to distribute a quantum key or a as memory for polarization entanglement in a quantum repeater architecture.

\begin{figure}[h!]
\includegraphics[width=\linewidth]{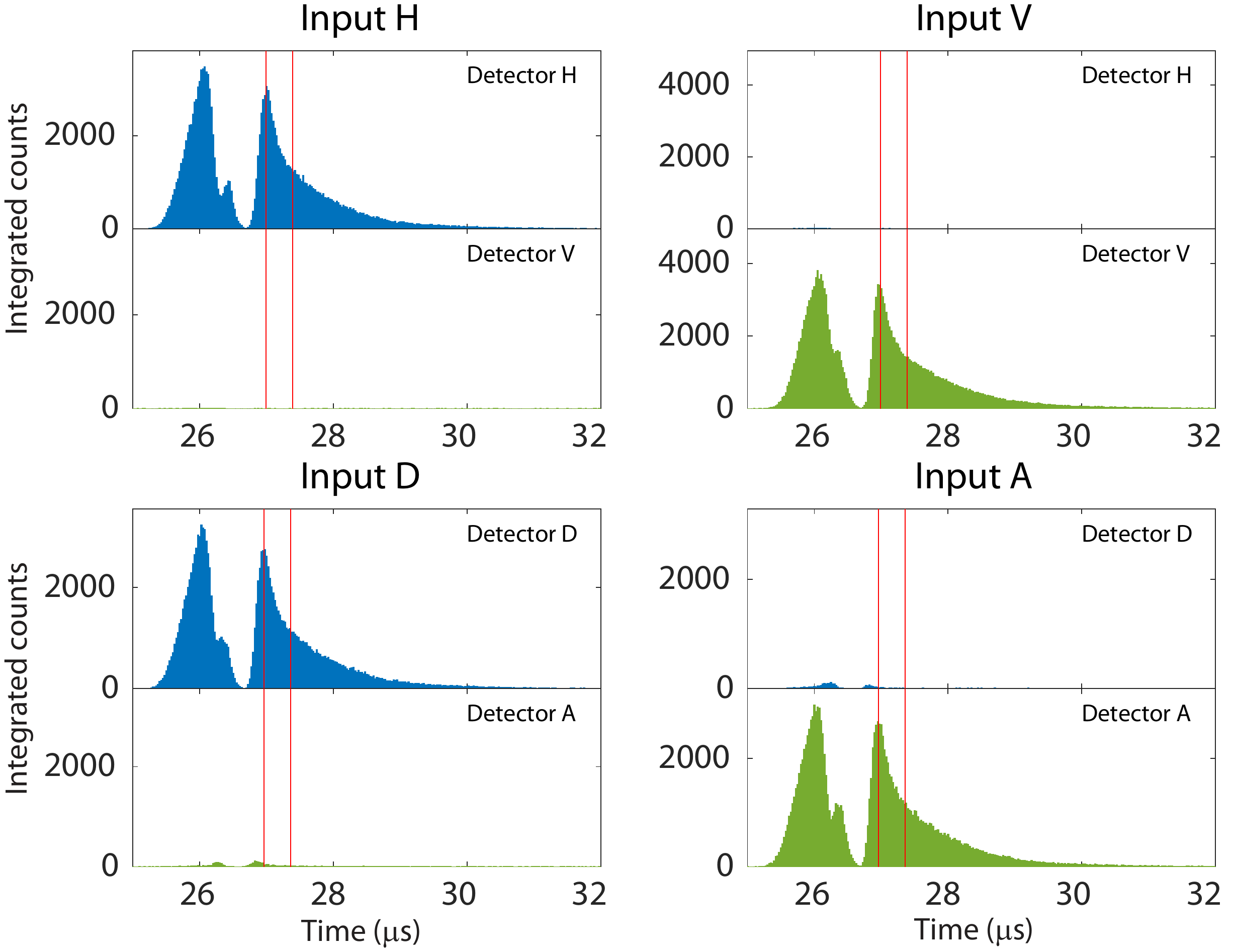}
\caption{\textbf{QBER evaluation of the long distance communication setup plus memory.} In the Bob site the polarization states are received and stored sequentially in a room temperature quantum memory. We randomly choose one of the Z and X bases to measure the polarization state, and then calculate the QBER over a region of interest equal to the input pulse width (red bars). We show histograms on the photons counts in each of the four polarizations. The first peak represents non-stored photon (leakage) while the second peak represents the retrieved photons. In an experiment with high input photon number, the obtained QBERs are less than 1\%, as it can be seen in the low counts corresponding to undesirable polarization detections.}
\end{figure}
\section{Experiment 2: Storage of a random sequence of polarization states with high photon number.}
After showing unconditional memory operation over the free space network, we now show that the network also operates with high fidelity on a pulse-by-pulse basis, demonstrated by full polarization analysis at Bob location. This is done by randomizing the polarization input of the experiment. Further insight into our current capabilities is obtained by analyzing the quantum bit error rates (QBER) $Q_X$ and $Q_Z$ for $X$ and $Z$ bases after propagation and storage. Starting with pulses containing high number of photons ($\sim$ 100 photons, see Fig. 3), we evaluated the QBER after storage of the random polarization states. An average QBER of 0.57\% for the two orthogonal bases have been measured within a region of interest equal to the input pulse width. This QBER is compatible with the typical error rate obtained in a standard quantum key distribution experiment. The importance of this result is two-folded: 1) the storage process at room temperature does not intrinsically add non-unitary rotation to the states, and in the limit of high signal-to-background has negligible effect on the total QBER; 2) the memory is capable of storing and retrieving a generic polarization qubits on a shot-by-shot level.

\section{Experiment 3: Storage of a random sequence of polarization qubits.}
In our next experiment the complete state measurement in the two bases was used again for an input of 1.6 photons before the memory, corresponding to 3.5 photons at Alice station. The evaluated QBERs after storage for polarization qubits are $Q_Z$= 11.0\% and $Q_X$ =12.9\% over a 100 ns region (see Fig. 4). The increase of the QBERs is only due to the background noise which is much more significant at the single-photon level. Nonetheless, the fidelities (corresponding to $1-$QBER) still remain higher than the classical limit for the corresponding storage efficiency. The latter result is rather counter-intuitive when dealing with superpositions $|D\rangle$ and $|A\rangle$ as it implies that the two rails forming the quantum memory store or miss the pulse coherently (in order to preserve the storage fidelity for that particular polarization), as opposed to retrieving rather $|H\rangle$ or $|V\rangle$ at any given time in a shot-by-shot experiment. This ability is crucial in networks performing quantum key distribution protocols and it also shows that the memory is currently capable of receiving entangled polarization states without distorting them. We do mention that this last experiment constitutes the quantum communication part of the well known BB84 protocol \cite{Benn84ieee}, with the addition of a synchronizing quantum memory between Alice and Bob.
\begin{figure}[h!]
\centering
\includegraphics[width=\linewidth]{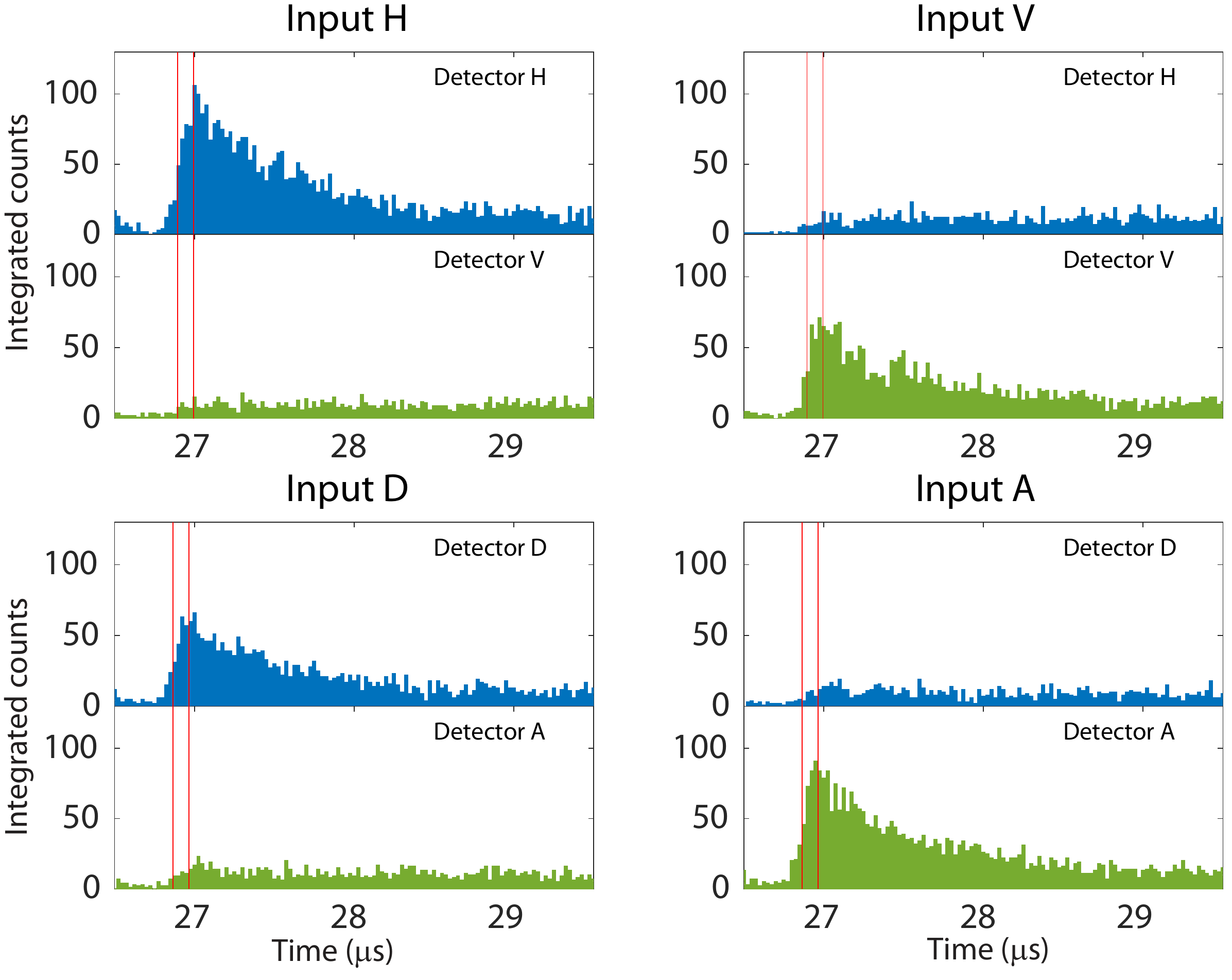}
\caption{\textbf{QBER evaluation for single photon level experiment.} (a) The QBER is calculated in a $100 ns$ window (red bars). QBER of  $11.0\%$ and $12.9\%$ are  respectively achieved for Z and X bases. At the single-photon level undesirable polarization rotations remain absent, noise in the orthogonal channel arises from control-field induced non-linear processes.}
\end{figure}

\section{Experiment 4: Noise-free operation and QBER improvements.}
In order to unlock the potential of our elementary realization as a quantum cryptography network, the main bottleneck identified in the aforementioned experiments is the memory performance at the required single-photon level. Naturally, there has to exist a compromise between room-temperature, all-environment operation and the background noise of the device at the quantum level, thereby creating limits to the achievable QBER. In our EIT configuration the single-photon level background mechanism is produced by several non-linear effects. Gearing towards increasing the performance of the memory we have performed experiments in which we have replaced one of the etalons in the filtering system with a similar unit with different free spectral ratio. This addition limits the background photons created by broad four-wave-mixing. To also eliminate the contribution due to incoherent scattering we have also added a weak ($\langle n \rangle \leq $ 0.01) re-pumping beam on resonance with the $5S_{1/2} F = 1$ $\rightarrow$ $5P_{1/2} F' = 1$ transition that remains on during the complete storage procedure. First results and our interpretation of the date indicate that a total suppression of the background noise can be achieved by engineering the interaction of the two created dark-state-polariton modes \cite{Karpa2008}.\\
Figure 5a shows the results of a one-rail experiment including the extra-repumper (light-blue). We can see that after retrieval, the two dark-state-polariton interaction creates regions without the additional background noise. In this experiment the re-pumper strength is increased to highlight the noise-free regions. We measured a SBR $\sim$ 26 for an input $\langle n \rangle \sim 1.3$ photons. We can then infer a corresponding fidelity of $97\%$ and QBER's $\sim3\%$ for $\langle n \rangle \sim 1$ (see caption in Fig. 5).

\begin{figure}[h!]
\includegraphics[width=\linewidth]{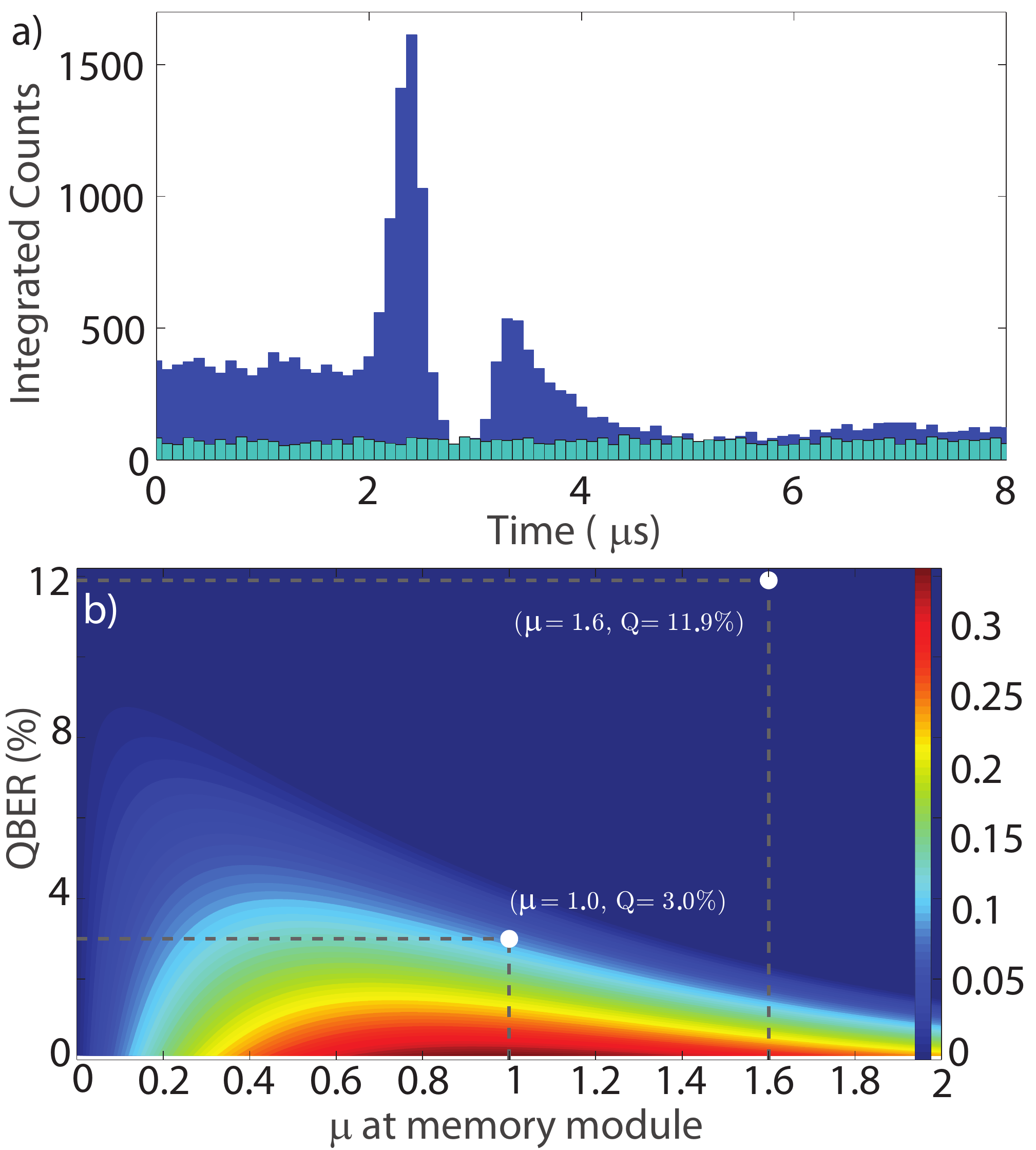}
\caption{\textbf{Noise-free quantum memory operation}. a) Noise reduction by introducing an auxiliary field (light-blue histogram), the interaction between dark-state-polaritons creates a background free region. Retrieving the probe under these conditions results in a SBR $>$25. The SBR is calculated using a 100 ns intergration region at the peak of retrieve signal and a minimized averaged background obtained in a 1 $\mu$s region centered around 5.2 us (divided by 10). (b) Quantum key distribution rate vs mean photon number and quantum bit error rate. Color bar represents the key rate. The line intersecting light blue and dark blue (negative key rate area) corresponds to the boundary for the positive key rate. The white dots indicate the regime of bare quantum memory and noise-free memory regimes.}
\end{figure}

The relevance of this new regime of operation is highlighted by analyzing its consequences in the achievable quantum key distribution rate (\emph{R}) per channel efficiency for sharing random secret key, encoded in random polarization states, between Alice and Bob. \emph{R} depends on the quantum bit error rate (QBER) and the mean photon number $\mu$. In the infinite key length limit, it is given by: $R=\mu ( e^{-\mu}(1-H(Q_X))-H(Q_Z)f(Q_Z))$, where $Q_X$ and $Q_Z$ are the QBERs, $H(x)$ is the binary Shannon entropy function and $f(Q_Z)$ is the efficiency of the classical error correction protocol. We have evaluated the absolute key rate vs. the input photon number and our average QBER with $f(Q_Z)$=1.05 \cite{elko11qic} for two cases. In the first case we include the bare quantum memory operation ($QBER=11.9\%$ for $\mu=1.6$). Fig. 5b shows that this regime lies just outside of the region for positive key rate generation, indicating not fully-secure qubit communication. This situation is fully-corrected by applying the noise reduction techniques explained above, as in this new regime the operation ($QBER=3\%$ for $\mu=1$) is well inside the secure communication threshold. This is a very important achievement as our quantum network has all the elementary capabilities for quantum cryptography operation.

\section{Experiment 5: Fully-portable quantum memory operation.}
In order to boost the achieved quantum network operation towards all-environment qubit connections between distant isolated locations, portable and robust quantum memories are paramount. In our last experiment we show the storage of single-photon level qubits in a first-prototype of a fully-portable plug-and-play memory. This prototype has the same features of the designs used in our aforementioned experiments but is fully independent of laboratory infrastructure as it only requires the probe photons and an EIT control field as inputs. It also possesses a miniaturized version of the filtering system with independent temperature controllers. A detailed depiction of the portable memory is shown in Fig. 6a. In Fig. 6b we show a storage of light experiment in which we store pulses with a mean photon number  $\langle n \rangle \sim 2$, in a single-rail experiment, corresponding to a SBR of 7.2.

\begin{figure}[h!]
\includegraphics[width=\linewidth]{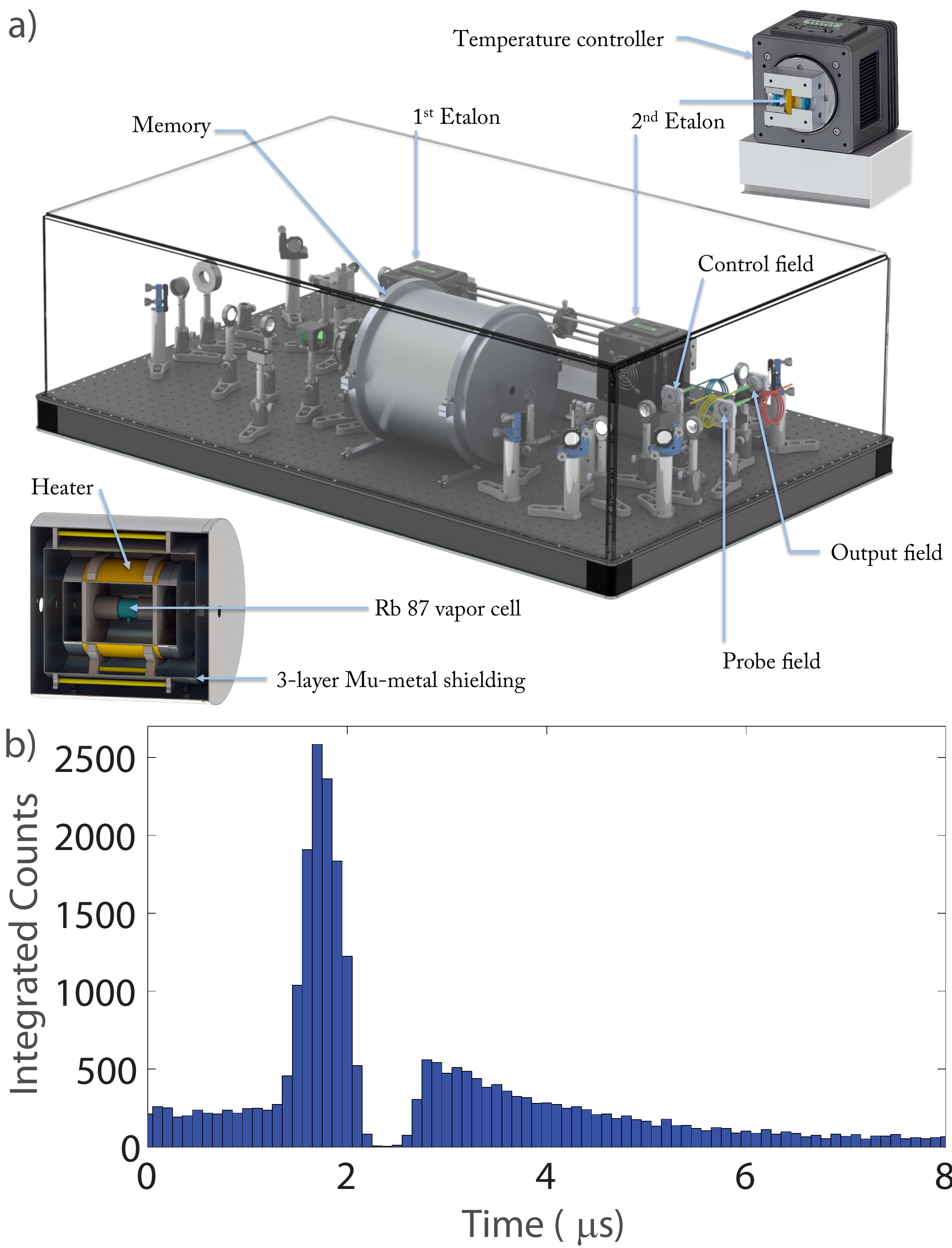}
\caption{a) Prototype of a room temperature portable quantum memory. Upper-right inset: Detail of one of the frequency filtering units, including the silica etalon, isolation oven, temperature control cold-plate and PID temperature regulation circuitry. Bottom-left inset: Detail of the interaction zone including the Rb cell, temperature control electronics and three-layer magnetic shielding.  b) Storage of single-photon level light pulses in the portable quantum memory. }
\end{figure}

\section{Conclusions.}
In conclusion, we have shown for a first time a network of quantum devices in which breakthrough operational capabilities are possible. We have achieved the first proof of principle combination of free-space propagation of random single-photon level polarization qubits and their storage and retrieval in a room temperature quantum memory.  These results effectively constitute the quantum part of the BB84 protocol with the addition of a quantum memory. Furthermore we have shown noise-suppression techniques that allow our network to operate in a regime useful for quantum cryptographic communication with low QBER's. Lastly, we have shown fully-portable room temperature quantum memory operation. Together, all these capabilities pave the way for more sophisticated applications using a network of portable quantum memories.\\
Because free-space propagation does not require the challenge of frequency conversion to the infrared, our setup can already be used for short-distance proof-of-concept memory-assisted device independent QKD experiments. For example, performing Hong-Ou-Mandel (HOM) photon interference using photons retrieved from two memories, together with memory-assisted temporal shaping of the outgoing qubits \cite{Namazi_cascading_2015,heinze_generation_2016,nisbet-jones_highly_2011} and applying the aforementioned noise-reduction techniques, will make it possible to store two random streams of qubits independently in each memory and to perform Bell measurements after simultaneous retrieval events. We also envisioned that the portable aspect of the memories will allow their use in remote observatory locations, opening a pathway for experiments with photons travelling over ultra-long satellite communication channels. The performance of our envisioned applications will benefit by a continuous development of our portable technology, including an increase in the speed (bandwidth) of the memory together with shorter pulse duration and an increase in the success rate of the storage procedure by means of heralding.
\section{Acknowledgments}
We thank Ruoxi Wang and Mael Flament for technical assistance in the development of the portable quantum memory setup. The work was supported by the US-Navy Office of Naval Research, grant number N00141410801, the National Science Foundation, grant number PHY-1404398 and the Simons Foundation, grant number SBF241180. B. J. acknowledges financial assistance of the National Research Foundation (NRF) of South Africa.


\end{document}